\documentclass[journal]{IEEEtran}
\usepackage{amsmath,amsfonts}
\usepackage{algorithm2e}
\usepackage{array}
\usepackage{textcomp}
\usepackage{stfloats}
\usepackage{url}
\usepackage{verbatim}
\usepackage{graphicx}
\usepackage{cite}
\usepackage{hyperref}
\usepackage{url}
\usepackage{booktabs}
\usepackage{braket} 
\usepackage[dvipsnames]{xcolor}
\usepackage{flushend}

\usepackage{xcolor}
\usepackage{soul}

\SetKwComment{Comment}{/* }{ */}
\RestyleAlgo{ruled}

\usepackage{xcolor} 

\usepackage{multicol}
\usepackage{multirow}

\usepackage{mwe}
\usepackage{float}
\usepackage{balance}
\usepackage{tikz}
\usetikzlibrary{quantikz}
\usepackage{xfrac}
\usepackage{amssymb}

\begin{document}
\title{Quanv4EO: Empowering Earth Observation by means of Quanvolutional Neural Networks}
\author{Alessandro Sebastianelli$^1$\thanks{$^1$ $\Phi$-lab, European Space Agency, Frascati, Italy. Email: alessandro.sebastianelli@esa.int, bls@ieee.org},~Francesco Mauro$^2$\thanks{$^2$Engineering Department, University of Sannio, Benevento, Italy. Email: \{f.mauro@studenti., ullo@\}unisannio.it},~Giulia Ciabatti$^3$\thanks{$^3$Department of Computer, Control and Management Engineering, Sapienza University of Rome, Rome, Italy. Email: giulia.ciabatti@uniroma1.it},~Dario Spiller$^4$\thanks{$^4$School of Aerospace Engineering, Sapienza University of Rome, Rome, Italy. Email dario.spiller@uniroma1.it},~\\~ Bertrand Le Saux$^1$,~Paolo Gamba$^5$\thanks{$^5$Department of Electrical, Computer and Biomedical Engineering, University of Pavia, Pavia, Italy. Email: paolo.gamba@unipv.it}~and Silvia Ullo$^4$}

\date{}
\maketitle
\markboth{Submitted to IEEE Transactions on Geoscience and Remote Sensing}{}

\begin{abstract}

A significant amount of remotely sensed data is generated daily by many Earth observation (EO) spaceborne and airborne sensors over different countries of our planet. Different applications use those data, such as natural hazard monitoring, global climate change, urban planning, and more.
Many challenges are brought by the use of these big data in the context of remote sensing applications. In recent years, employment of machine learning (ML) and deep learning (DL)-based algorithms have allowed a more efficient use of these data but the issues in managing, processing, and efficiently exploiting them have even increased since classical computers have reached their limits. 
This article highlights a significant shift towards leveraging quantum computing techniques in processing large volumes of remote sensing data. The proposed Quanv4EO model introduces  a quanvolution method for preprocessing multi-dimensional EO data. First its effectiveness is demonstrated through image classification tasks on MNIST and Fashion MNIST datasets, and later on, its capabilities on remote sensing image classification and filtering are shown.
Key findings suggest that the proposed model not only maintains high precision in image classification but also shows improvements of around 5\% in EO use cases compared to classical approaches. Moreover, the proposed framework stands out for its reduced parameter size and the absence of training quantum kernels, enabling better scalability for processing massive datasets.
These advancements underscore the promising potential of quantum computing in addressing the limitations of classical algorithms in remote sensing applications, offering a more efficient and effective alternative for image data classification and analysis.

\end{abstract}


\section{Introduction}

\IEEEPARstart {M} {any} challenges emerge with the use of big data in the context of remote sensing (RS) applications, and even if the employment of machine learning (ML) and deep learning (DL)-based algorithms have allowed more efficient use of these data, the issues in managing, processing, and efficiently exploiting them have even increased since classical computers have reached their limits \cite{ChiMin}, \cite{articleclassicalvsquantum}.

{Q}{uantum} Convolutional or Quanvolutional Neural Networks (QuanvNNs) represent an exciting discovery and development in the field of DL, where the principles of quantum computing (QC) are integrated into convolutional neural networks (CNNs). QuanvNNs aim to enhance the data processing and representation in the computer vision and Earth observation (EO) domain by introducing quantum-inspired operations.

At the core of QuanvNNs are the quanvolutional operations, which replace or augment the traditional convolutions used in classical CNNs. Quanvolutions leverage quantum-inspired filters that capture complex quantum states, allowing for more expressive feature representations. These filters can be implemented using various quantum-inspired techniques such as quantum circuits, quantum gates, or other quantum-inspired operations. By using these quantum filters, QuanvNNs can learn to extract features that exploit the inherent quantum nature of the data, leading to improved discrimination and generalization capabilities in various computer vision tasks \cite{henderson_quanvolutional_2019}.

This article introduces Quanv4EO, a quanvolution method for preprocessing multi-dimensional EO data. 
The work presented represents a significant milestone in the field of EO by introducing the application of QCNNs to the field of RS and paving the way for overcoming current limitations of classical computing and further enhancing the use of quantum computing for EO tasks.
One hurdle in utilizing quantum computing for EO tasks is related to the requirement for large quantum circuits to process the sheer size of EO images.
To harness the full potential of quantum computing in this domain, the proposed approach provides a promising solution to handle large RS images with remarkable effectiveness.

With the proposed method, only a few qubits are needed to encode the information, rather than hundreds of them, while performances are maintained to the same level (effectively, they are improved) with respect to comparable state-of-the-art approaches.

QCNNs, as already mentioned, incorporate quantum-inspired operations into the architecture of CNNs, which have demonstrated in the last years to draw new paths for EO image processing and classification. The proposed solution opens up new avenues for leveraging the capabilities of QC in many EO tasks, such as image classification, object detection, and semantic segmentation.

Differently from previous works by the same authors \cite{zaidenberg2021advantages, sebastianelli2021circuit}, where a hybrid quantum neural network with a classical convolutional branch and a quantum layer as classifier was proposed, in this paper the quantum layers have been moved at the beginning, while the classification has been left to classical strategies. Moreover, kernel functions in high dimensional Hilbert spaces have been accessed. \\
Overall, the quantum advantage of QCNNs is demonstrated with applications specifically to RS and EO domains. Moreover, a lazy training-like procedure is proposed: a method of learning wherein generalization from the training data is postponed until a query is directed to the system. All the above characteristics make our model completely new and able to overcome some of the highlighted challenges. As discussed ahead, the main objective of our work is  not necessarily to assert superior performance but rather to demonstrate that the proposed approach could potentially serve as a resolution to challenges encountered in fully quantum approaches.

The paper is organized as follows. The state-of-the-art (SOTA) is analyzed in Section~\ref{sec:related_works} by introducing related works on QuanvNNs. The quanvolution approach and the related kernel functions are introduced in Section~\ref{sec:methodology}. In Subsection~\ref{sec:lazy} the lazy training-like procedure proposed for our model is explained.
In order to reduce the simulation time, a parallelized version of the quantum convolution operation is developed, as explained in Subsection~\ref{sec:parallelization}.
In Subsection~\ref{sec:time} the processing time is evaluated by varying a list of certain parameters, such as the size of the input image, the stride, the number of qubits, the kernel size, and the number of output channels. 
In Subsection~\ref{sec:feature_maps}, the evaluation of the feature maps, produced by a quanvolutional layer by varying the number of qubits and kernel size, is presented.

In Section~\ref{sec:experimental_results} the quality and the usability of features extracted by these layers is evaluated, by validating our method through image classification by using both common datasets for image classification in the data science field and EO datasets (Subsections~\ref{results_datascience} and~\ref{results_on_EO}).
Comparisons carried on with the classical counterparts demonstrate how the feature maps extracted by using quantum kernels increase the performance of our method. Extensive analysis is also performed, highlighting how the main objective of our work is not necessarily to assert superior performance but rather to demonstrate that the proposed approach could reach these performance with a significantly reduced number of parameters. Discussions and conclusions, reported in Section~\ref{sec:disc_conc}, close the paper.

\section{Related works on Quanvolution} \label{sec:related_works}
Convolutional neural networks (CNNs) have rapidly gained popularity in numerous machine learning applications, especially within the domain of image recognition. Much of the significant advantage offered by these networks stems from their capacity to extract features from data in a hierarchical fashion. These features are obtained through various transformational layers, with the convolutional layer being a prominent component that lends the model its name.

QuanvNNs were first introduced in \cite{henderson2020quanvolutional}, where transformational layer known as a ``quantum convolution" or ``quanvolutional" layer were proposed. Quanvolutional layers process input data by locally transforming it through a series of random quantum circuits, similar in principle to the transformations performed by randomly applied convolutional filter layers. Assuming that these quantum transformations yield meaningful features suitable for classification, the overall algorithm may prove highly valuable in the context of near-term quantum computing. This is because it necessitates the use of compact quantum circuits with minimal or no error correction.\\
Specifically, in the quanvolutional approach, akin to a traditional convolutional layer, the quanvolutional layer generates feature maps by applying localized transformations to input tensors. However, instead of conducting element-wise matrix-matrix multiplications, the quanvolutional layer initiates the process by encoding an image patch into a quantum state. Subsequently, this state undergoes a sequence of operations involving two-qubit gates, like the CNOT gate, and parameterized one-qubit gates, such as rotations around the three axes of the Bloch sphere \cite{mattern2021variational}. Given the constraints on the quantum hardware's size, processing an entire image is impractical. A work around this limitation, is the use of quanvolution with smaller patches, as employed in convolution, using the available qubit count to encode the information. Importantly, the encoding of compact filter sizes does not necessitate Quantum Random Access Memory (QRAM)   
\footnote{QRAM is necessary for algorithms that require efficient access to large amounts of quantum data, such as the Grover's algorithm, which benefits from quick access to a quantum database for searching a specific element.
The main limitation of QRAMs is storage capacity, which is influenced by the complexity and scalability of quantum systems. Not having QRAM can be seen as positive in certain contexts because it avoids the technical challenges and limitations associated with their implementation. QRAMs are susceptible to decoherence and noise, which can make scaling them up difficult. Additionally, the need for external control for each QRAM query can introduce a significant opportunity cost, where the control hardware could be more efficiently used to run extremely parallel classical algorithms \cite{bugalho2023resource}.} technology, which remains under development.
QuanvNNs have a remarkable ability to discern and understand localized patterns within data, which is crucial for various tasks such as image recognition or signal processing. This capability stems from their inherent design, which allows them to effectively capture and represent intricate details present in different regions of an image. Moreover, QuanvNNs maintain translational invariance across the entire image, meaning that they can recognize patterns regardless of their position within the image. This characteristic is essential as it ensures that the network's learning remains consistent irrespective of shifts or transformations in the input data.
Furthermore, the integration of quanvolutional layers into modern layered machine learning models marks a significant advancement in leveraging quantum technology for practical applications. These quanvolutional layers seamlessly blend into existing architectures, facilitating the adoption of quantum techniques without requiring a complete overhaul of established frameworks. 

Building upon these advancements, the research reported in \cite{mattern2021variational} delves into the exploration of quantum image encoding techniques and their impact on the performance of a hybrid quanvolutional algorithm inspired by convolutional methods. The results indicate that the choice of image encoding is important,  that there is no one-size-fits-all solution, and the final selection depends on the specific application.

Since the majority of current quantum circuits operates on one-dimensional objects, when dealing with two-dimensional data, like images, to encode them in quantum states, they must be transformed  into a one-dimensional array. In \cite{onuoha2022data}, the authors present various techniques for mapping 2D data to a 1D format. These techniques are explored within the domain of quanvolutional neural networks.
Every patch of an input image holds pixel data that is employed for quanvolutional processing in QCNN, akin to the traditional CNN's convolution. The pixel arrangements within each patch may be altered into various scanning sequences before being inputted into the quantum encoder for the extraction of features.
Experimental results reveal that the suggested mapping methods deliver substantial improvements in comparison to the conventional raster scan approach. 

A QuanvNN  extends a CNN by incorporating quanvolutional layers with quantum circuits for filters, showing potential benefits. In the use-case presented in \cite{sooksatra2021evaluating}, a convolutional layer was replaced with a quanvolutional layer, resulting in similar classification accuracy but reduced loss. Models with quanvolutional layers resisted adversarial examples and were resilient to Circuit-Weighted Adversarial (CWA) attacks. 

In a Quanvolution algorithm, every pixel's data is encoded within individual qubits. Once feature extraction is performed, a feature corresponding to the measurement outcome of each qubit in the Parameterized Quantum Circuit (PQC) is derived. It is important to note that the quantity of features obtained through this process is equivalent to the number of qubits in the PQC, which coincides with the size of the kernel.
In \cite{baek2022scalable}, a scaled QCNN (sQCNN) was introduced to enhance the number of filters. Since an increased number of qubits in a single quantum circuit often leads to barren plateaus, the objective of sQCNN is to mitigate barren plateaus by augmenting the number of filters, which in turn increases the number of quantum circuits rather than the qubits within the circuit. The sQCNN employs multiple filters, thereby providing flexibility in adjusting the quantity of extracted features.

It is important to highlight that Quanvolutional layers leverage quantum circuits in the creation of feature maps. The research proposed in \cite{patil2022implementation} utilizes random quantum circuits as filters. A comparison between QuanvNN and CNN models can help determine whether the addition of a quanvolutional layer enhances model accuracy and performance. Both models were trained using various training dataset sizes. Beyond dataset size, the model was also tested with different numbers of Quanvolutional layers (1, 4, 8, 16, 32) using 600 training images and 150 test images. The model's performance improves as the number of layers increases, but this also leads to increased computational time. Similarly, in \cite{cong2019quantum} QuanvNNs using random quantum circuits was introduced for better feature extractions using kernels in higher dimensions.

\section{Methodology} \label{sec:methodology}
\begin{figure}[!ht]
	\centering
	\includegraphics[width=1\columnwidth]{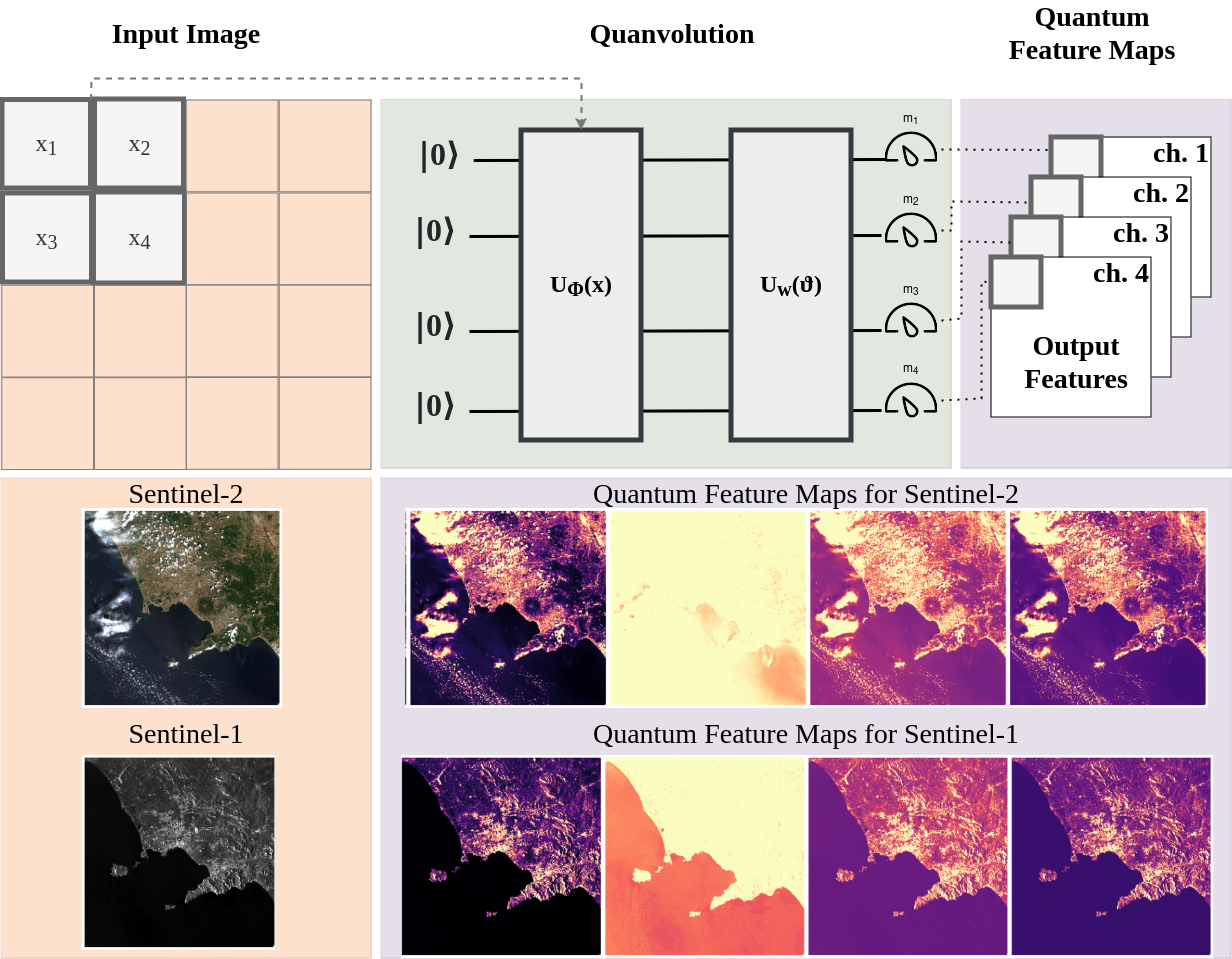}
	\caption{Quanvolution Scheme, where is shown a quantum kernel of $2\times2$ qubits (top). Features extracted with the quantum kernel on a Sentinel-1 and Sentinel-2 images (bottom).}
	\label{fig:qconv}
\end{figure}

A possible approach and description for quantum convolution (Quanvolution) directly derives from its classical counterpart. The establishing procedure is schematized in Fig. \ref{fig:qconv}, where for simplicity we show the quantum convolution with a kernel of size $2\times 2$ and a quantum circuit of $4$ qubits.

Let's assume that the quantum circuit used for the quanvolution operator is the Variational Quantum Circuit (VQC) in Figure \ref{fig:qconv}. With this configuration, a small region of the input image defined by the $kernel\_size$ is encoded in a quantum circuit of at least $kernel\_size^2$ qubits to the quantum state of Hilbert $\mathbb{H}^{2^n}$ through the operator $U_{\phi}(x)$. The ansatz $U_W(\vartheta)$, with random parameters $\vartheta$ and gates ($R_x$, $R_y$, $R_z$ and CNOT) \cite{henderson2020quanvolutional}, is applied to the quantum state $\ket{\phi(x)}$ \cite{de2024towards}. Measurements operations are done to get the output, the expected values with respect to the observable $\hat{O}$ are calculated, and the results are obtained as:
\begin{equation}
	\centering\label{eqn:fxt}
	f(x,\vartheta) = \bra{\phi(x)}U^{\dagger}_{W}(\vartheta)\hat{O}U_W(\vartheta)\ket{\phi(x)}
\end{equation}

Under the conditions of ``lazy training regime", as explained later in Section \ref{sec:lazy}, the parameters $\vartheta$ are frozen, but ultimately they can be optimized using a common gradient-descent technique:
\begin{equation}
	\centering\label{eqn:upd}
	\nabla_{\vartheta}(f(x,\vartheta)) = \frac{1}{2}\left[f(x,\vartheta + \frac{\pi}{2}) - f(x,\vartheta - \frac{\pi}{2}) \right]
\end{equation}

this aspect will not be treated in this paper and it is left for future works.

The mathematical formulation of quanvolution follows the one of the convolution in signal theory. Let assume that $X \in \mathbb{R}^{W\times H}$ is a single-band image of spatial dimension $W$ and $H$, width and height respectively and $k \in \mathbb{R}^{M\times N}$ a kernel with dimension $M$ and $N$, where typically  $ M = N $ and $M \ll W$ and $N \ll H$. The convolution of $X$ and $k$ is given by Equation \eqref{eqn:con}.

\begin{equation}
	\centering
	\label{eqn:con}
    \begin{split}
	y[m,n] &= X[m,n] * k[m,n] = \\&= \sum_{i=-\infty}^{+\infty}\sum_{j=-\infty}^{+\infty} X[i,j] \cdot k[m-i, n -j]
    \end{split}
\end{equation}

considering that both $X$ and $k$ are finite signals, the equation can be rewritten as for Equation \eqref{eqn:con2}

\begin{equation}
	\centering
	\label{eqn:con2}
	y[m,n]  = \sum_{i=-m}^{m+W}\sum_{j=-n}^{n+H} X[i,j] \cdot k[m-i, n -j]
\end{equation}

for the case of quanvolution Equation \eqref{eqn:con2} can be rewritten as for  Equation \eqref{eqn:con3}

\begin{equation}
	\centering
	\label{eqn:con3}
	\begin{split}
	y[m,n]  &= \sum_{i=-m}^{m+W}\sum_{j=-n}^{n+H} f(X[i,j], k[m-i, n -j]) \\
		   &= \sum_{i=-m}^{m+W}\sum_{j=-n}^{n+H} \bra{\phi(X[i,j])}U^{\dagger}_{W}(k[m-i, n-j])\\&\hat{O}U_W(k[m-i, n-j])\ket{\phi(X[i,j])}
	\end{split}
\end{equation}

The introduced formulation assumes that the input image has only one channel and the convolution is 2D. Yet, this assumption can be easily extended to the case of multi-channels image with 2D convolution (each channel is processed by a 2D kernel and results are combined properly), as well as multi-channels image with 3D convolution (the kernel is 3D, so the channels are processed together).

Moreover, also in the case of quantum convolution the stride has the same meaning with respect to its classical counterpart, unlike the depth of the feature map that for quantum convolution is directly linked to the number of qubits, since it cannot surpass the number of available qubits.



\subsection{Lazy Training Regime}\label{sec:lazy}
By definition, ``Lazy Learning'' or ``Lazy Training'' includes those ML algorithms where the generalization of the training data is delayed until a query to the system is made, as opposed to ``Eager Learning'' or ``Feature Training'' algorithms, where a generalization is attempted before receiving a query and some features are learned \cite{Geiger_2020}, \cite{chizat2020lazy}. This approach proves to be particularly useful when the training data are continuously updated with new samples and in the case of large, and continuously changing training sets with few attributes that are commonly queried. A typical example of a lazy classifier is the K-Nearest Neighbors - KNN - algorithm. Specifically, the target function is approximated locally: this allows to solve multiple problems and deal with changes in the problem domain. In the case of deep learning, when dealing with the lazy-training regime, the dynamics are almost linear, and the Neural Tangent Kernel - NTK - almost does not change after initialization. In fact, in the NTK limit, the dynamics become linear with respect to the weight changes, hence it can be described by a frozen kernel $\Theta$. In the NTK limit, the learning dynamics are described as: 

\begin{equation}
    \centering
    \Theta(w, x_1, x_2) = \nabla_w f(w, x_1) \cdot \nabla_w f(w, x_2)
\end{equation}
which is the NTK definition and where: $x_1, x_2$ are the inputs, and $\nabla_w$ is the gradient with respect to the parameters $w$. 
Defining with $h$ the number of hidden neurons in the neural network, the $\Theta$ kernel evolves in time, but, as $h \rightarrow \infty$, $\Theta(w, x_1, x_2) \rightarrow \Theta_\infty (x_1, x_2)$. Hence, the kernel is frozen and the dynamics converge on a time independent of $h$ to a global loss minimum. 

The Lazy Training Regime has been discussed and tested also in the case of QML \cite{henderson2020quanvolutional}, \cite{PRXQuantum.3.030323}. In particular, in \cite{PRXQuantum.3.030323} a version of \textit{Quantum} Neural Tangent Kernel - QNTK - is defined and its dynamics in the frozen limit, where variational angles do not change much - i.e. in lazy-training regime - are analytically solved.

In \cite{henderson2020quanvolutional}, a more practical application of lazy training for the quantum part of the QuanvNN architecture is discussed. Here, a quantum transformation layer is added to a classical CNN architecture, enabling the embedding of a set of quanvolutional filters within the network's architecture. Notably, these quanvolutional filters utilize random quantum circuits. Their unique characteristic lies in their ability to efficiently access kernel functions in high-dimensional Hilbert spaces. This lazy training regime is adopted for several reasons. Firstly, it is derived from the insights presented in \cite{henderson2020quanvolutional}. Additionally, our experiments have yielded promising results even without training the quantum layers. Furthermore, the optimization of quantum convolution for EO use cases remains a time-consuming task, although ongoing research efforts, as discussed in \cite{ceschini2024}, are directed towards addressing this challenge.

\subsection{Parallelization and Optimization} \label{sec:parallelization}
One of the main challenges for quantum convolution consists of the processing time. Contrary to its classical counterpart (e.g. convolutional layer implemented in TensorFlow, PyTorch, etc.) that has been developed and optimized for years, to the best of authors knowledge, there is not an optimized version of the quantum convolution. In order to mitigate this problem we developed a parallelized version of the quantum convolution operation.

To show the validity of the method, and the advantage introduced in our implementation through parallelization, in this section we present several experiments both on different hardware and on different testing conditions. The two experimental hardware used are the following:

\begin{itemize}
    \item \textbf{Setup-A}: Ubuntu, CPU: 11th Fen Intel(R) Core(TM) i7-11800H @ 2.30GHz, CORES: 8, LOGICAL PROCESSORS: 16, RAM: 32.0 GB, GPU: NVIDIA GeForce RTX 3080 16GB of dedicated RAM.
    \item \textbf{Setup-B}: MacOS, CPU: Intel Core i5 @ 2.4 GHz, CORES: 4, LOGICAL PROCESSORS: 8, RAM: 8.0 GB, GPU: Intel Iris Plus Graphics 655 1536 MB of dedicated RAM.
\end{itemize}

To evaluate the processing time of the quantum convolution, we used an image of dimension $(64,64,3)$.

In Table \ref{tab:time} we report the processing times by comparing the two experimental setups and quantum convolution with and without parallelization. As expected, Table \ref{tab:time} clearly shows that each configuration with parallelized quantum convolution has a lower processing time, the best results are obtained with Setup-A when using 16 logic processors.

\begin{table}[!ht]
    \centering
    \large
    \caption{Processing time for Quanvolution}\label{tab:time}
    \resizebox{0.65\columnwidth}{!}{\begin{tabular}{lcc}
    \toprule
    Modality & \multicolumn{2}{c}{Processing Time (s)}\\
    \midrule
    
                 & Setup-A & Setup-B\\
    \cmidrule(r){2-3}
    
    No Para.             & $\sim 5.72$ & $\sim 9.3$\\
    Para. 4-logi. proc.  & $\sim 1.54$ & $\sim 4.4$\\
    Para. 8-logi. proc.  & $\sim 1.10$ & $\sim 3.2$\\
    Para. 16-logi. proc. & $\sim 0.95$ & -\\
    \bottomrule
    \end{tabular}}
\end{table}

Parallelization has been achieved through the Python library \href{https://joblib.readthedocs.io/en/stable/}{joblib}, and the implementation can be found in \href{ttps://github.com/alessandrosebastianelli/quanvolutional4eo}{GitHub - main branch}.
Further, we have managed to reach a much lower processing time (as better described in Section \ref{sec:time}) by developing an additional version using the Jax library, that can be found in \href{https://github.com/alessandrosebastianelli/quanvolutional4eo/tree/dev-as-jax-v2}{GitHub - jax branch}.

\subsection{Processing time analysis}\label{sec:time}
In this section, we properly evaluated the processing time by varying some parameters, such as the size of the input image, the stride, the number of qubits, the kernel size, and the number of output channels.

\begin{figure*}[!ht]
    \centering
    \includegraphics[width=0.8\columnwidth]{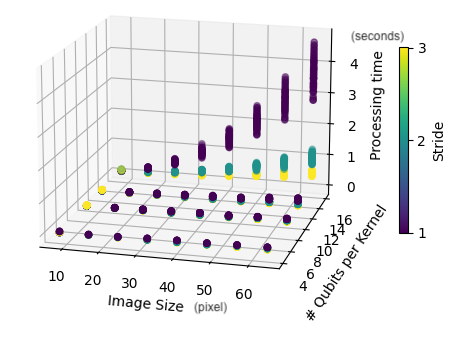}~\includegraphics[width=0.8\columnwidth]{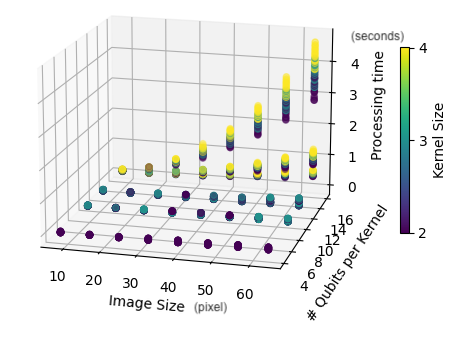}~\\
    \includegraphics[width=0.8\columnwidth]{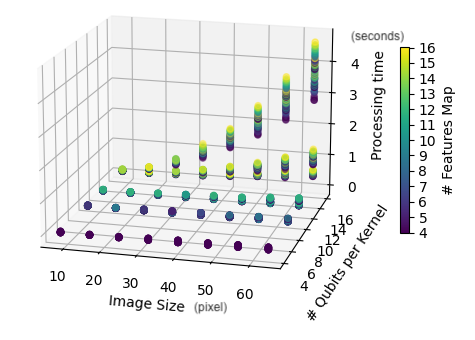}
    \caption{Quanvolution Image Processing Time by varying size of the input image, stride, number of qubits, kernel size and number of output channels. The top left plot shows how the timing is influenced by number of qubits image size and stride. The top right shows how the timing is influenced by number of qubits, image size and kernel size. The bottom plot shows how the timing is influenced by the number of qubits, image size and number of output feature maps.}
    \label{fig:timing}
\end{figure*}

\begin{figure}[!ht]
    \centering
    \includegraphics[width=\columnwidth]{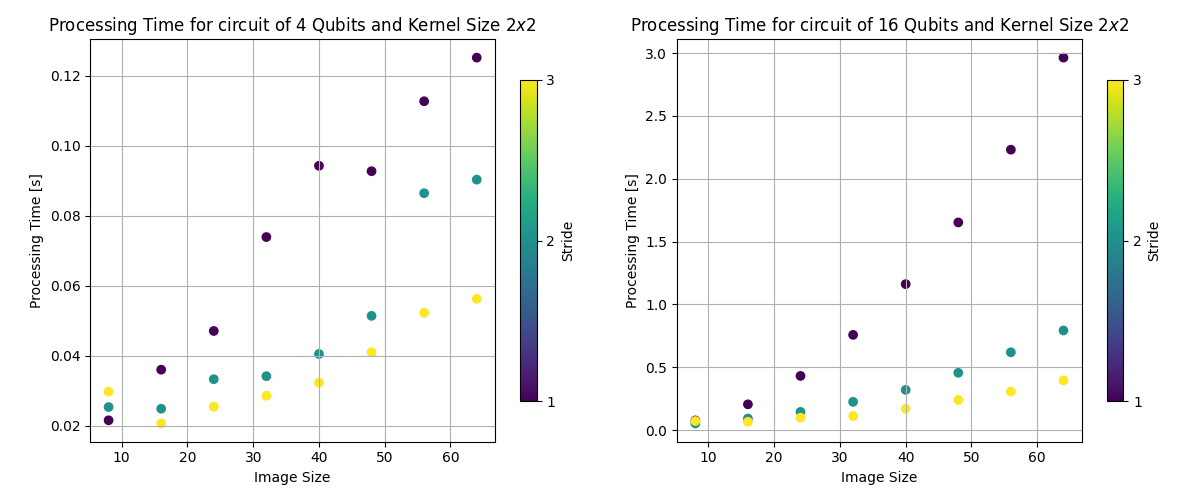}
    \caption{Quanvolution processing time with a configuration with 8/16 qubits, $2\times 2$ kernel size and 4 output features map.}
    \label{fig:timing2}
\end{figure}

As Figure \ref{fig:timing} shows, the processing time is influenced by image size, indeed the processing time slowly increases with the increase of the size of the input image. The number of qubits also influences the time required by the quanvolutional layer to process one image, and this, in combination with the image size and stride, is more evident at the bottom of the Fig. \ref{fig:timing}. The stride has a huge influence on the processing time, indeed from the bottom plot we can see that with the increase of the stride size (from deep blue to yellow), the processing time drastically decreases. This result was expected since a larger stride results in less application of the quantum circuit and so fewer operations necessary to process one image. The Kernel size is also impacting on the processing time, but less than the stride, in fact the top right plot of Fig. \ref{fig:timing} shows that with the increase of the kernel size (from deep blue to yellow) the processing time gets worse. This result was also expected since the kernel size is strictly related to the number of qubits. The same happens for the number of the output feature maps: the bottom plot shows that increasing the number of output feature maps results in higher processing time, which again relates to the number of qubits. For simplicity, breaking apart graphs in Fig. \ref{fig:timing}, by selecting experiments with 8 and 16 qubits, a kernel size of $2\times 2$, and a number of output features maps equal to 4, it is possible to have a clearer view of processing time (depending on the size of the image) as shown in Fig. \ref{fig:timing2}. 

\subsection{Feature Maps Consistency} \label{sec:feature_maps}
In this section, we evaluate the feature maps produced by a quanvolutional layer by varying the number of qubits and kernel size.

\begin{figure*}[!ht]
    \centering
    \resizebox{2\columnwidth}{!}{
    \begin{tabular}{cc}
    \resizebox{0.1\columnwidth}{!}{Kernel Size} &  \\
    \toprule
    \resizebox{0.05\columnwidth}{!}{$4\times 4$} & \includegraphics[width=\columnwidth]{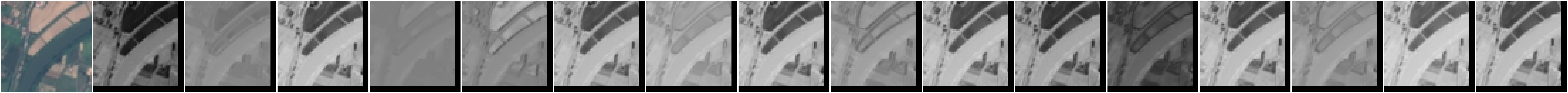}\\ 
    \resizebox{0.05\columnwidth}{!}{$3\times 3$} & \includegraphics[width=\columnwidth]{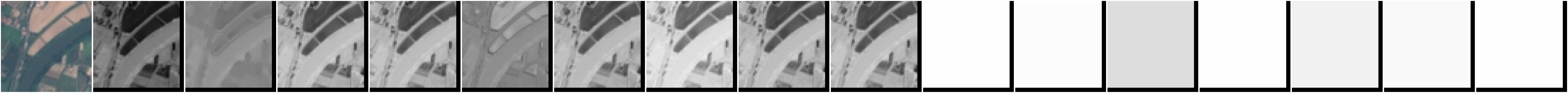}\\ 
    \resizebox{0.05\columnwidth}{!}{$2\times 2$} & \includegraphics[width=\columnwidth]{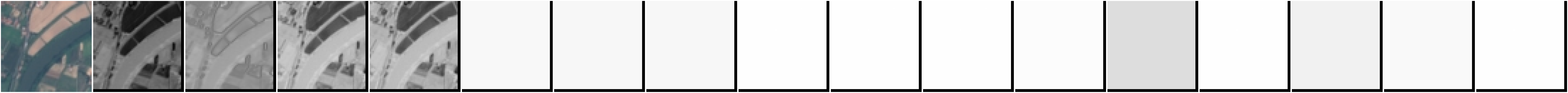}\\ 
    \end{tabular}}
    \caption{Features Map with random circuit of 16 qubits}
    \label{fig:ft16}
\end{figure*}

\begin{figure*}[!ht]
    \centering
    \resizebox{2\columnwidth}{!}{
    \begin{tabular}{cc}
    \resizebox{0.1\columnwidth}{!}{Kernel Size} &  \\
    \toprule
    \resizebox{0.05\columnwidth}{!}{$3\times 3$} & \includegraphics[width=\columnwidth]{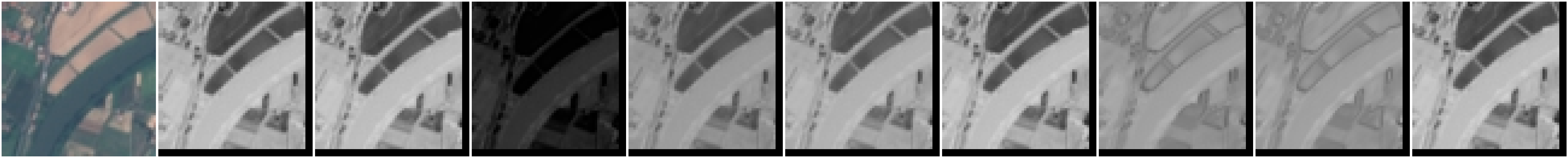}\\ 
    \resizebox{0.05\columnwidth}{!}{$2\times 2$} & \includegraphics[width=\columnwidth]{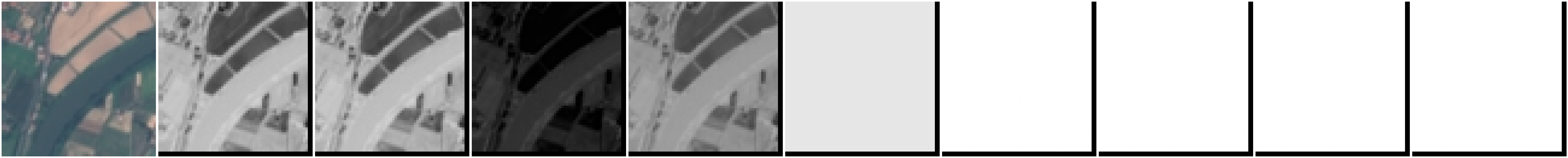}\\ 
    \end{tabular}}
    \caption{Features Map with random circuit of 9 qubits}
    \label{fig:ft9}
\end{figure*}

\paragraph{Random Circuit} Given the structure of quanvolutional with random layer, as proposed by Henderson \textit{et Al.}\cite{henderson2020quanvolutional}, there is a connection between the kernel size and the number of features maps that can be produced. Indeed, as show by Fig. \ref{fig:ft16} and Fig. \ref{fig:ft9}, when we request $\# features > (kernel\_size)^2$ we only get $(kernel\_size)^2$ informative feature maps. This depends on the fact that a random layer is not well connecting all the qubits, so only the qubits used to ingest the image pixels are well activated to produce the feature maps.

\begin{figure*}[!ht]
    \centering
    \resizebox{2\columnwidth}{!}{
    \begin{tabular}{cc}
    \resizebox{0.1\columnwidth}{!}{Kernel Size} &  \\
    \toprule
    \resizebox{0.05\columnwidth}{!}{$3\times 3$} & \includegraphics[width=\columnwidth]{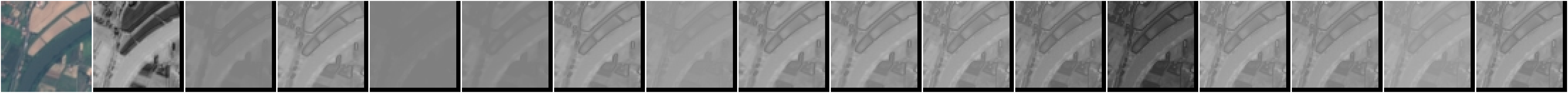}\\ 
    \resizebox{0.05\columnwidth}{!}{$2\times 2$} & \includegraphics[width=\columnwidth]{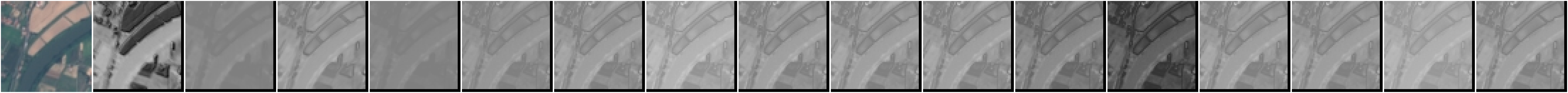}\\ 
    \end{tabular}}
    \caption{Features Map with custom random circuit of 16 qubits}
    \label{fig:ft16n}
\end{figure*}
\begin{figure*}[!ht]
    \centering
    \resizebox{2\columnwidth}{!}{
    \begin{tabular}{cc}
    \resizebox{0.05\columnwidth}{!}{Layer} &  \\
    \toprule
    \resizebox{0.1\columnwidth}{!}{quanvolution} & \includegraphics[width=\columnwidth]{imgs/highlyentangled-q16-k3.png}\\ 
    \resizebox{0.05\columnwidth}{!}{CNN}          & \includegraphics[width=\columnwidth]{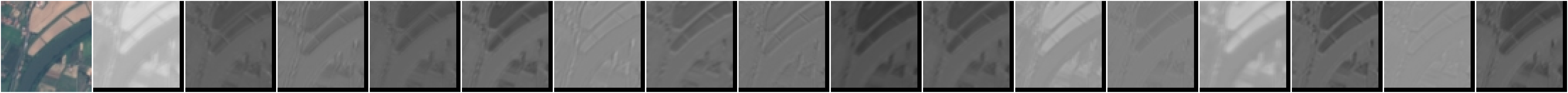}\\ 
    \end{tabular}}
    \caption{Quanvolutional VS Convolutional Features Maps}
    \label{fig:quanvVScnn}
\end{figure*}

\paragraph{Custom Circuit} A solution for the above-mentioned problem can be the implementation of circuit with a custom component on top of the random layer. For example by mutually connecting all the qubits with CNOT gates and then by applying a random layer, it is possible to get 16 features maps even with a kernel size of $3\times 3$ or $2\times 2$, as shown in Fig. \ref{fig:ft16n}. 

\paragraph{Differences with Convolutional Layers} A comparison between quantum convolutional and classical convolutional feature maps is presented in Fig. \ref{fig:quanvVScnn}, showcasing qualitative differences (they have the same scale). Notably, despite lacking training, quantum feature maps appear to contain more information than their classical counterparts. This enhancement is especially evident in classical feature maps with minimal information, indicating the effectiveness of the quantum approach. Additionally, the use of \href{https://docs.pennylane.ai/en/stable/code/api/pennylane.RandomLayers.html}{random circuits} draws a parallel with classical weight initialization, underscoring the distinction in initialization methods between classical and quantum contexts (on a side the weights and on the other the gates).

\section{Experimental Results} \label{sec:experimental_results}

\begin{figure*}
    \centering
    \includegraphics[width=2\columnwidth]{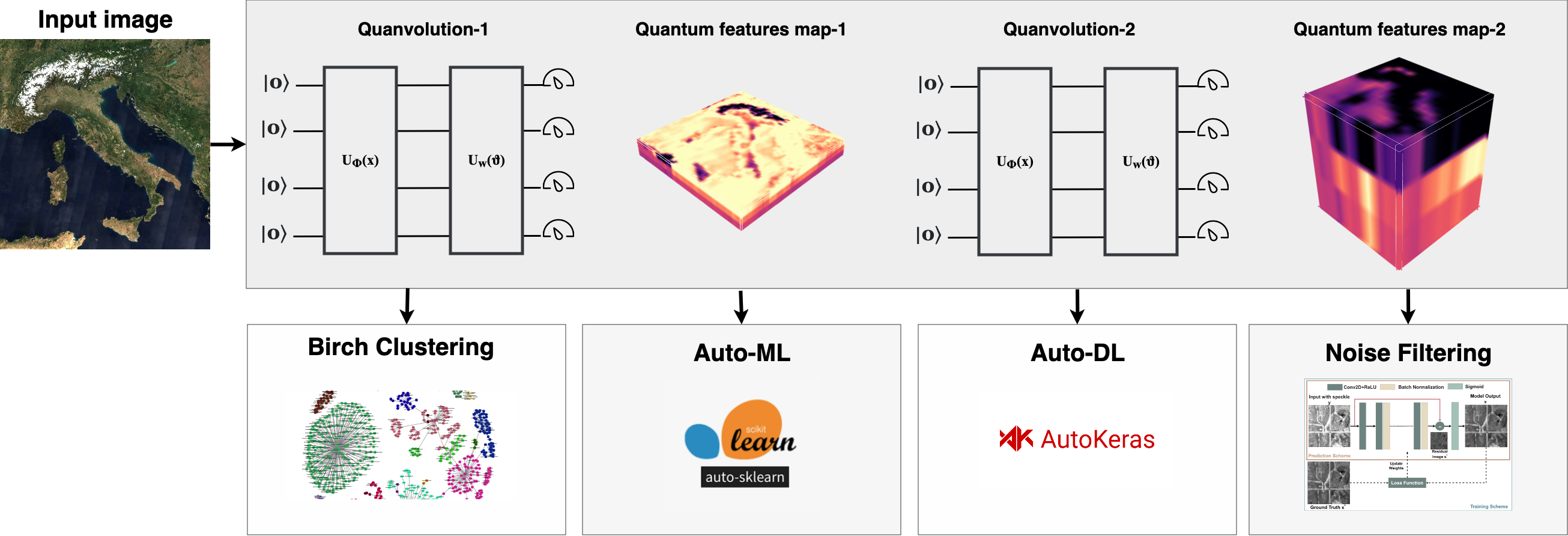}
    \caption{Schematich of the proposed QuanvNN for EO data classification. On the left site the two quanvolutional layers and on the right side the three classical AI methods.}
    \label{fig:quanvolution-classification}
\end{figure*}

In quantum machine learning (QML), the training process often involves optimizing quantum circuits or variational parameters representing the model. Training a quantum model necessitates manipulation of quantum states, measurements, and optimization algorithms, tasks that can be computationally expensive and resource-intensive. Our quanvolutional layers, essentially fixed quantum kernels processing EO data, aim to efficiently access kernel functions in high-dimensional Hilbert space. It's crucial to note that while QML can exploit quantum system properties for efficient computations, direct manipulation of a "huge Hilbert space" in terms of individual quantum states is not typical. Quantum computing operates within a Hilbert space, representing a quantum system's state space, with its size exponentially growing with qubit count, a factor promising for specific computations. QML algorithms often focus on quantum variational algorithms (e.g., Variational Quantum Eigensolver, Quantum Approximate Optimization Algorithm) or quantum neural networks, utilizing quantum circuits for tasks like optimization or classification, leveraging quantum mechanics principles. Despite their potential, QML algorithms still face limitations and are in early development stages. Directly accessing and manipulating a "huge Hilbert space" typically necessitates a fully realized, fault-tolerant quantum computer, a long-term goal in quantum computing research \cite{schuld2021machine}. \\

To assess feature quality and usability extracted by these layers, we conducted several experiments, firstly on data science datasets and then on Earth Observation datasets. Regading the former, we ran MNIST and Fashion MNSIT classification. For the latter, we ran land use and land cover classification experiments within a simulated \textit{Pennylane} environment, utilizing various methodologies, as illustrated in Figure \ref{fig:quanvolution-classification}: 1) Balanced Iterative Reducing and Clustering using Hierarchies (BIRCH) clustering, an unsupervised method, 2) Automated Machine Learning (autoML), a supervised ML method, and 3) Automated Deep Learning (autoDL), a supervised DL method. Moreover, in another case study, we demonstrated the efficiency of QCNNs in extracting speckle noise from Synthetic Aperture Radar (SAR) data.


Our  work builds off \cite{wilson2019quantum}, where classical data are processed by means of quantum circuits, and the generated output - as ``quantum" feature maps - is then passed as input to train a linear model. In this paper, first of all, a more complex NN is utilized - a CNN-like model. Moreover,  the quantum layer is not trained, but it only serves as a feature extractor and, thus, is kept ``frozen". This  approach is classified as  a lazy training-like procedure.

\subsection{Resuls on Data Science datasets} \label{results_datascience}
In order to validate the proposed method we initially ran some experiments using non EO dataset, particularly the MNIST \cite{lecun1998gradient} and the Fashion MNIST dataset \cite{xiao2017fashion}.\\

\paragraph{\textbf{MNIST dataset}}
The results obtained by the Quan4EO method on the MNIST dataset demonstrate high accuracy, reaching $0.9984$, with a remarkably small model size of only $42k$ parameters plus $16$ qubits (frozen). This positions our method in competition with the most advanced models in image classification on MNIST dataseet, such as Branching CNN + HVC, EnsNet, and Efficient-CaspNet. Despite its small size (significantly reduced compared to other models in the SOTA), our method maintains high precision, offering an efficient and promising alternative for image data classification. 

\begin{table}[!ht]
    \centering
    \caption{Comparisons with SOTA on MINST data classification. \href{https://paperswithcode.com/sota/image-classification-on-mnist}{(Classical Model rank)}} \label{tab:results_mnist}
    \begin{tabular}{lcc}
    \toprule
    Model & Accuracy & Model Size\\
    \midrule
    Branching CNN + HVC \cite{byerly2021no}       & 0.9987 & 1.5 M\\
    EnsNet \cite{hirata2023ensemble}              & 0.9984 & - \\
    Efficient-CaspNet \cite{mazzia2021efficient}  & 0.9984 & 160 k\\
    SOPCNN \cite{assiri2020stochastic}            & 0.9983 & 1.4 M\\
    RMDL \cite{kowsari2018rmdl}                   & 0.9982 & -\\
    \midrule
    Quanv4EO                                      & 0.9984 & 42 k + 16q (frozen)\\
    \bottomrule
    \end{tabular}
\end{table}

\paragraph{\textbf{FashionMNIST dataset}}
Similarly, on FashionMNIST dataset, the Quan4EO method proved to be highly competitive, achieving an accuracy of $96.81\%$. Furthermore, the model remains incredibly small, with the same size as the model used for MNIST classification ($42 k + 16$ qubits, frozen). This result is really encouraging, since our method offers an excellent balance between model size and performance.

\begin{table}[!ht]
    \centering
    \caption{Comparisons with SOTA on Fashion MNIST data classification. \href{https://paperswithcode.com/sota/image-classification-on-fashion-mnist}{(Classical Model rank)}} \label{tab:results_fmnist}
    \begin{tabular}{lcc}
    \toprule
    Model & Accuracy & Model Size\\
    \midrule
    Fine-Tuning DARTS \cite{tanveer2021fine}      & 0.9691 & 3.9 M\\
    Shake-Shake \cite{foret2020sharpness}         & 0.9641 & - \\
    PreAct-ResNet18 + FMix \cite{harris2020fmix}  & 0.9636 & -\\
    Random Erasing \cite{zhong2020random}         & 0.9635 & -\\
    E2E-3M \cite{phong2020rethinking}             & 0.9592 & -\\
    \midrule
    Quanv4EO                                      & 0.9681 & 42 k + 16q (frozen)\\
    \bottomrule
    \end{tabular}
\end{table}

In both cases, the results confirm the effectiveness and efficiency of the Quan4EO method in image data classification, whether on standard datasets like MNIST and FashionMNIST or on Earth Observation (EO) datasets, as previously demonstrated. Our technique provides a solid foundation for application across a wide range of contexts, ensuring high-level performance with reduced computational requirements.

\subsection{Results on Earth observation datasets} \label{results_on_EO}

\paragraph{Features Clustering}

This section presents the clustering results when using the unsupervised Birch Clustering to the feature maps extracted using Quanvolution. Moreover, we repeated the same experiments, under the same settings, using instead feature maps extracted with classical convolution.

BIRCH Clustering is a hierarchical clustering algorithm designed for large-scale datasets. It aims to efficiently and effectively cluster data by incrementally building a clustering feature tree. BIRCH algorithm clusters data points by maintaining a compact summary of the dataset, called the Cluster Feature (CF) tree, which contains information about the clustering structure. The algorithm utilizes a combination of distance metrics and density-based clustering to determine cluster centers and assign data points to clusters. BIRCH is particularly suited for applications with large datasets and provides a scalable approach to clustering \cite{zhang1997birch}.

Before presenting the results, it is worth to highlight that BIRCH Clustering is not meant to work with images. Our adaptation to images worked smoothly. Nevertheless, our scope was to test Quanvolutional features maps using as many learning strategies as possible, thus leading us to the choice of using this kind of unsupervised strategy also in this case.

In order to work with a data format as close as possible to the one more compatible with BIRCH Clustering, we applied 2 quanvolutional layers with a kernel size of 4 and stride 4 on the EuroSAT dataset \cite{helber2019eurosat} obtaining for each image in the dataset a feature map of $4\time 4\times12$. We then applied the Principal Component Analysis (PCA)  to compress the dimension of the features from 192 ($4\time 4\times12$) to 10. We then applied BIRCH Clustering to this new set of variables. 

\begin{figure}[!ht]
    \centering
    \begin{tabular}{c}
    \includegraphics[width=\columnwidth]{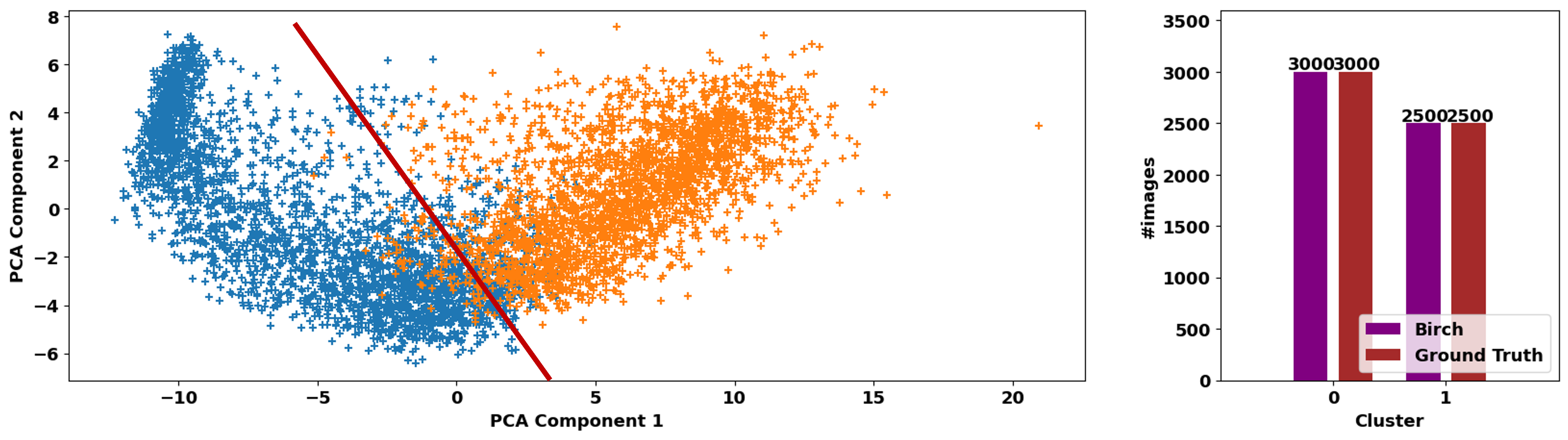} \\
    \includegraphics[width=\columnwidth]{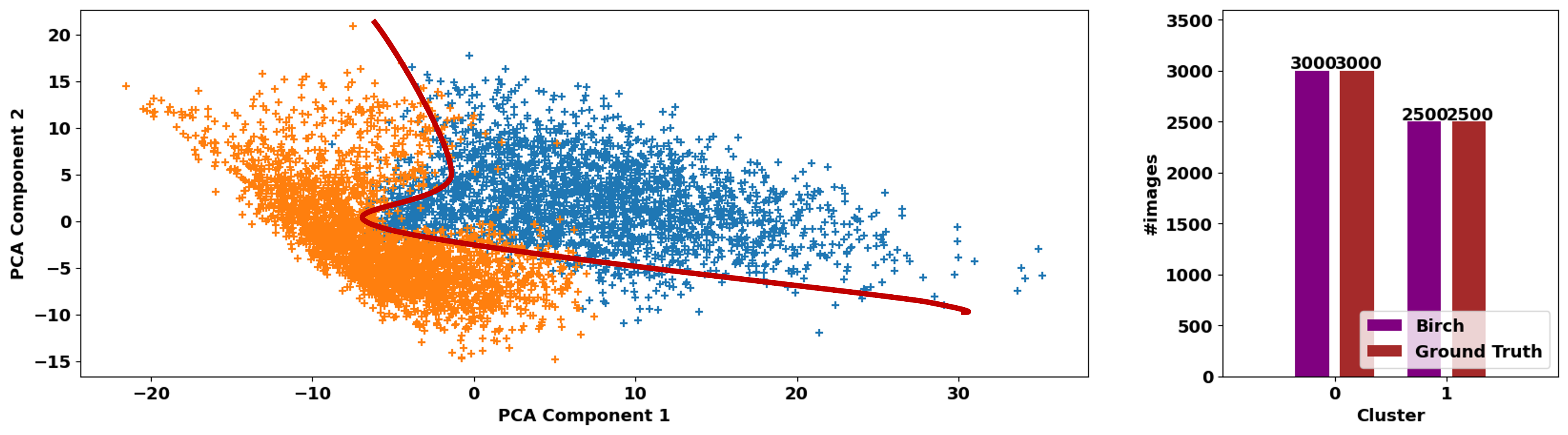}
    \end{tabular}
    \caption{Features clustering with Quanvolution (top) and with CNN (bottom) - Binary Case.}
    \label{fig:quanvclusters}
\end{figure}

\begin{figure}[!ht]
    \centering
    \begin{tabular}{c}
    \includegraphics[width=\columnwidth]{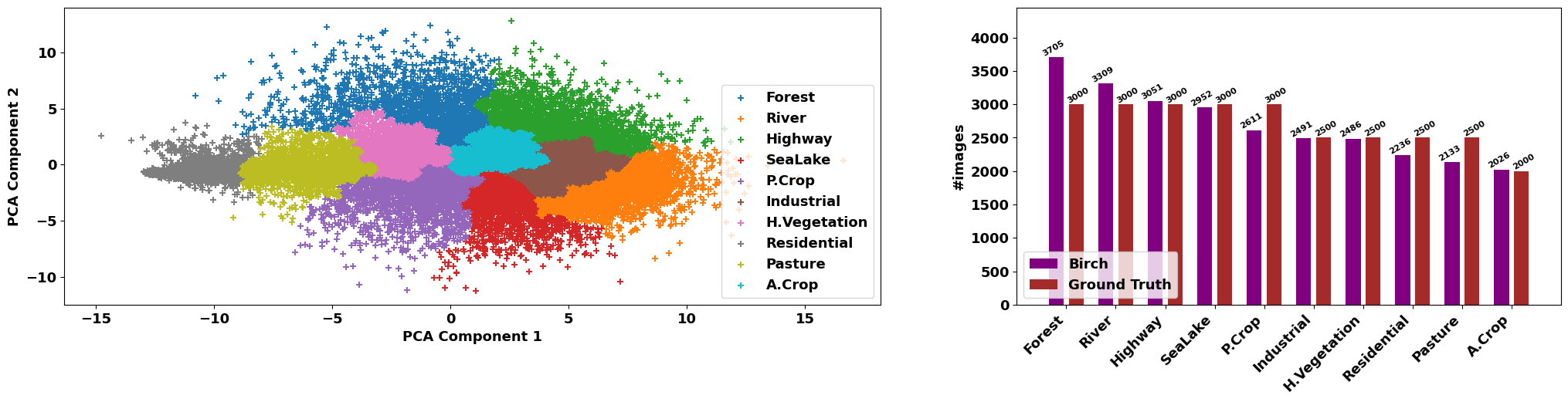}\\
    \includegraphics[width=\columnwidth]{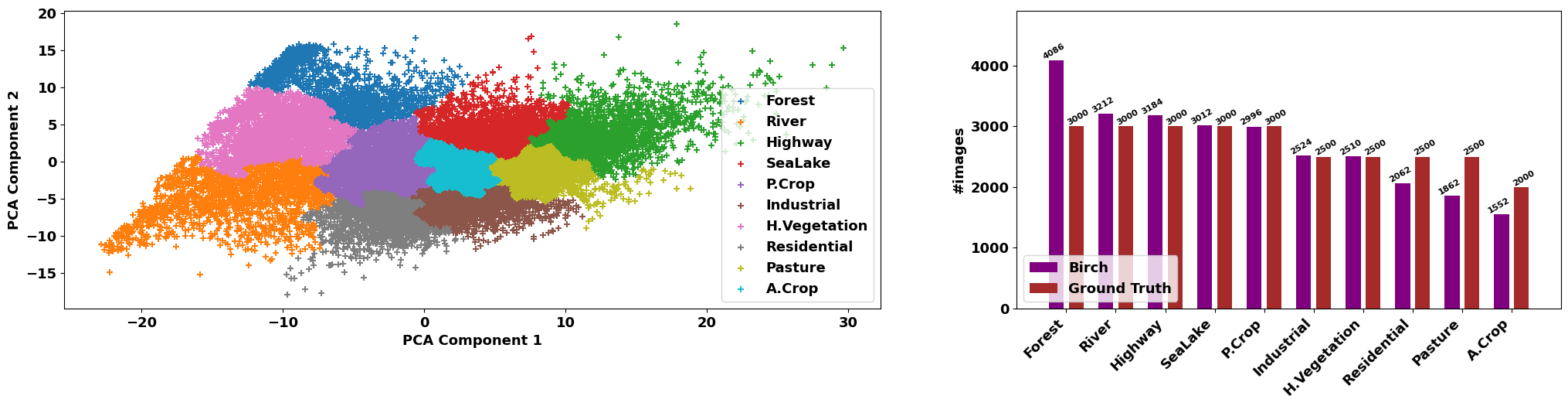}
    \end{tabular}
    \caption{Features clustering with Quanvolution (top) and with CNN (bottom) - Multiclass Case}
    \label{fig:quanvclusters_all}
\end{figure}

In the end, we made plots to show that quanvolution could better differentiate  the features, as shown at the top of Figure \ref{fig:quanvclusters}. Indeed, they are more distant with respect to the CNN ones shown at the bottom of Figure \ref{fig:quanvclusters}. Even if the clustering worked perfectly on both methods, the quanvolution  works required a simpler model, indeed drawing a straight line is sufficient to split the features with an acceptable error, while for the convolutional one a polynomial function is needed. 
In a first phase (simple scenario) we applied clustering on only 2 classes of our dataset (Figure \ref{fig:quanvclusters}) and then (more complex scenarios) we expanded the analysis to the whole dataset (Figure \ref{fig:quanvclusters_all}). Since EuroSAT is a labeled dataset we can compute the quality of clustering by counting the items clustered in the right class. We found that, in the simple scenario both models achieved a score of 100\%, but the quantum model is better in making the features more differentiable; in the more complex scenario, quanvolution produced 8\% miss-classified labels, while the simple convolution produced 11\% miss-classified labels. 

Since the dataset is labeled, although we are working with an unsupervised method, it is still possible to calculate classification (clustering) accuracy. Results are reported in Tab. \ref{tab:res_clustering}, where we compare the results of the same BIRCH clustering model trained on Quanvolution feature maps and on classic features maps, respectively. Results show that the latter performed better in most of the classes, apart from River and Permanent Crop class where the quanvolutional features maps are beaten by the classic ones, while for the class Sea Lake, both models have shown the same performance. Generally speaking, the feature maps extracted using a quantum kernel introduce a $2\%$ of gain. 

\begin{table}[!ht]
    \centering
    \caption{Results of Birch Clustering for QCNN and CNN feature maps.\\}\label{tab:res_clustering}
    \begin{tabular}{l|cc}
         \toprule
         Class          & CNN & QCNN\\
         \midrule
         Annual Crop    & 0.87      & \bf{0.99} \\
         Forest         & 0.85      & \bf{0.89} \\
         H. Vegetation  & 0.99      & \bf{0.99} \\
         Highway        & 0.97      & \bf{0.99} \\
         Industrial     & 0.99      & \bf{0.99} \\
         Pasture        & 0.85      & \bf{0.92} \\
         P. Crop        & \bf{0.99} & 0.93      \\
         Residential    & 0.90      & \bf{0.94} \\
         River          & \bf{0.97} & 0.95      \\
         Sea Lake       & \bf{0.99} & \bf{0.99} \\
         \midrule
         Mean Acc.      & 0.94      & \bf{0.96} \\
         \bottomrule
    \end{tabular}
\end{table}

\paragraph{AutoML Classifier with quantum features}

This section presents the classification results when using the AutoML strategy to the feature maps extracted using Quanvolution. Moreover, we repeated the same experiments, under the same settings, using feature maps extracted with classical convolution.

AutoML is a process that automates the development of ML models. It simplifies tasks such as data pre-processing, feature engineering, algorithm selection, hyperparameter tuning, and model evaluation. AutoML leverages AI and ML techniques to automatically search for the best model and configuration. Overall, AutoML accelerates model development, enables non-experts to use ML, and improves efficiency and effectiveness in building models \cite{hutter2019automated}. Moreover, AutoML helps to reduce/remove the model-selection bias that can be introduced by researchers.

Before presenting the results, it is worth highlighting that AutoML is designed for tabular data. Although we successfully adapted it for images, this adaptation might affect the results. In any case, since the goal was to test Quanvolution feature maps with a variety of learning strategies, the choice was to use also the AutoML.

Experiments have been run using the \textit{AutoSklearnClassifier} from \textit{Auto Sci-Kit learn} Python library. The full configuration  can be found on our Git-Hub page [github citation],  briefly allowing the \textit{AutoSklearnClassifier} to search in the list of classifiers (including Multilayer Perceptron, Random Forest, etc.) for the one with the best accuracy score, without constructing an ensemble of models. The time for the search is set to 1200 seconds in total, dedicating 120 seconds to each model with a memory limit of 8GB.

Results are reported in Tab. \ref{tab:res_automl}, where we compare the results of the same AutoML model trained on Quanvolution feature maps and on Classic features maps respectively. Results show that the former performed better in most of the classes, apart from Highway and Permanent Crop class where both models performed the same. Generally speaking, the feature maps extracted using a quantum kernel introduced $2\%$ of gain. 

\begin{table}[!ht]
    \centering
    \caption{Results of AutoML applied to QCNN and CNN feature maps.\\}
    \label{tab:res_automl}
    \begin{tabular}{l|cc}
         \toprule
         Class          & CNN & QCNN\\
         \midrule
         Annual Crop    &     0.90  & \bf{0.91} \\
         Forest         &     0.96  & \bf{0.98} \\
         H. Vegetation  &     0.84  & \bf{0.85} \\
         Highway        & \bf{0.85} & \bf{0.85} \\
         Industrial     &     0.93  & \bf{0.95} \\
         Pasture        &     0.89  & \bf{0.91} \\
         P. Crop        & \bf{0.85} & \bf{0.85} \\
         Residential    &     0.90  & \bf{0.93} \\
         River          &     0.87  & \bf{0.89} \\
         Sea Lake       &     0.95  & \bf{0.96} \\
         \midrule
         Mean Acc.      &     0.89  & \bf{0.91} \\
         \bottomrule
    \end{tabular}
\end{table}

\paragraph{AutoDL classifier with quantum features}

As for the previous section, this one presents the results when AutoDL is applied on quanvolution and classic convolution feature maps. Similarly to AutoML, which automates the development of ML models, AutoDL automates the development of deep learning models. It simplifies tasks such as architecture design, hyperparameter tuning, and model evaluation. AutoDL utilizes algorithms and techniques to automatically search for the best DL model, saving time and effort. By automating the process, AutoDL enables users to focus on problem formulation and interpretation of results. It accelerates the DL model development and it makes it more accessible \cite{jin2019auto}, and also in this case, it helps to reduce/remove the model-selection bias that can be introduced by researchers.

Experiments have been run using the \textit{AutoKeras Classifier} from \textit{AutoKeras} Python library. The full configuration used can be found on our Git-Hub page briefly allowing the \textit{AutoKeras Classifier} to search for the best \textit{Vannilla} classifier (simple CNNs), that is  the one with the best accuracy score, without constructing an ensemble of models. 

Results are reported in Tab. \ref{tab:res_autodl}, where we compare the results of the same AutoDL model trained on Quanvolution feature maps and on Classic features maps respectively. Results showed that the former performed better for all the classes. The feature maps extracted using a quantum kernel have introduced a $5\%$ of gain. 
\begin{table}[!ht]
    \centering
    \caption{Results of AutoDL applied to QCNN and CNN feature maps.\\}
    \label{tab:res_autodl}
    \begin{tabular}{l|cc}
         \toprule
         Class          & CNN  & QCNN\\
         \midrule
         Annual Crop    & 0.88 & \bf{0.93} \\
         Forest         & 0.94 & \bf{0.98} \\
         H. Vegetation  & 0.80 & \bf{0.87} \\
         Highway        & 0.81 & \bf{0.88} \\
         Industrial     & 0.92 & \bf{0.95} \\
         Pasture        & 0.83 & \bf{0.91} \\
         P. Crop        & 0.77 & \bf{0.88} \\
         Residential    & 0.94 & \bf{0.97} \\
         River          & 0.85 & \bf{0.90} \\
         Sea Lake       & 0.96 & \bf{0.98} \\
         \midrule
         Mean Acc.      & 0.88 & \bf{0.93} \\
         \bottomrule
    \end{tabular}
\end{table}

In summary, by means of quanvolution we were able to reach better performance, with an improvement of 2-5\% with respect to standard CNN. It can be observed that the BIRCH clustering is working the best, and this is amenable to the fact that hyperparameter selection for this algorithm is much easier than AutoML and AutoDL. Indeed, it is worth highlighting that for both cases, AutoML and AutoDL, we had to limit the search to a certain amount of time. Moreover, we did not explore all the possibilities of these two tools. Given that, it is possible, with more time and extended exploration, to reach better performances.

The important finding relies on the fact that in all cases we get an improvement, and specifically for the case of AutoDL, which reached better performances for all the classes. This is magnified by the fact that quanvolutional layers are ``frozen", hinting that results can be even better. Moreover, we used only two layers of quanvolution, so the global model is not that complex. This finding is in line with \cite{sebastianelli2021circuit}, indeed this paper demonstrates that hybrid-quantum solution can easily achieve better/comparable performance with respect to classical solutions, but with a light-weight structure.

Moreover we also observed that quantum models have a faster convergence - less number of epochs to reach the convergence - with respect to standard models, of 80\% on average. This finding is also in line with other works in the same field \cite{sebastianelli2021circuit, ceschini2022hybrid,sebastianelli2023quantum,incudini2023resource}.

Table \ref{tab:comparisons_SOTA} poses a challenge in explanation. In this study,  quantum processing is strategically positioned at the network's outset, delegating the classification task to clustering (unsupervised ML), autoML, and auto DL. The intention is not necessarily to assert superior performance but rather to sufficiently demonstrate that the proposed approach could potentially serve as a resolution to challenges encountered in fully quantum approaches. In future works, this can be reached by integrating a quantum classifier \cite{sebastianelli2021circuit} following the quanvolution layers, thereby avoiding significant issues tied to machine size when applied to Earth Observation (EO) data.

Furthermore, as we continue to optimize the model, there is the prospect of training the quanvolution filters, which is anticipated to yield even more favorable results. The presented table indicates that the proposed solutions exhibit comparable performance metrics, emphasizing the simplicity of the models—particularly evident in their compact sizes. Notably, the proposed approach outperforms specific models introduced in \cite{sebastianelli2021circuit} (Ry Circuit, Bellman Circuit, and Real Amplitudes circuit), showcasing superiority akin to the Coarse-to-fine grain method outlined in the same reference. Additionally, the obtained results surpass those achieved by some classical models.

\begin{table}[!ht]
    \centering
    \caption{Comparisons with SOTA on EuroSAT dataset classification}\label{tab:comparisons_SOTA}
    \resizebox{\columnwidth}{!}{
    \begin{tabular}{lcc}
        \toprule
        Model & Overall Accuracy & Model Size\\
        \midrule
        Helber et Al. \cite{helber2019eurosat} ResNet-50     & $0.98$ &   $25$ M\\
        Helber et Al. \cite{helber2019eurosat} GoogleNet     & $0.98$ &    $7$ M\\
        Li et Al. \cite{li2020deep} ResNet-18                & $0.98$ &   $11$ M\\
        Sumbul et Al. \cite{sumbul2019bigearthnet}           & $0.70$ &   $23$ k\\
        \midrule
        Sebastianelli et Al. \cite{sebastianelli2021circuit} (Ry Circuit)           & $0.79$ &   $42k + 4q$\\
        Sebastianelli et Al. \cite{sebastianelli2021circuit} (Bellman Circuit)      & $0.84$ &   $42k + 4q$\\
        Sebastianelli et Al. \cite{sebastianelli2021circuit} (Real Amplitudes)      & $0.92$ &   $42k + 4q$\\
        Sebastianelli et Al. \cite{sebastianelli2021circuit} (Coarse-to-fine grain) & $0.97$ &   $4\times(42k + 4q)$\\
        \midrule
        Quanv4Eo + Clustering                            & $0.96$ &    $16q$ (frozen)\\
        Quanv4Eo + AutoML                                & $0.91$ &    $16q$ (frozen)\\
        Quanv4Eo + AutoDL                                & $0.93$ &    $42k+16q$ (frozen)\\
        \bottomrule
    \end{tabular}}
\end{table}

\paragraph{QSpeckleFilter}
Observing Earth through (SAR) provides valuable insights regardless of weather conditions. However, SAR imagery often suffers from speckle noise, hindering accurate interpretation. As an application of QCNNs, in \cite{mauro2024qspecklefilter} a QSpeckleFilter is proposed, which outperforms its classical counterpart (based on CNNs), demonstrating promising advancements in EO applications.
 QSpeckleFilter builds upon a previous work by Sebastianelli et al. \cite{sebastianelli2022speckle}, utilizing quantum convolution and a modified speckle denoiser. The proposed method significantly enhances performance metrics, indicating the effectiveness of quanvolution in SAR image processing.
By processing the dataset with Quanvolutions, we expanded the dataset dimensions, enhancing noise extraction capabilities.
Following preprocessing using quanvolution, the architecture proposed in \cite{sebastianelli2022speckle} has been fine-tuned to incorporate quanvolution-preprocessed data.
The proposed method utilizes quanvoluted feature maps to estimate the speckle noise within each image. Subsequently, the identified speckle is subtracted from the Sentinel-1 image using a \textit{skip connection} and a \textit{subtraction layer}, resulting in a residual model. The skip connection serves as a bypass for the CNN, while the subtraction layer performs mathematical subtractions between the input data propagated through the skip connection and the CNN output.
The methodology demonstrates improved Peak Signal-to-Noise Ratio (PSNR) and Structural Similarity Index Measure (SSIM) values compared to previous method, as shown in Table \ref{QSpeckle}.
\begin{table}[!ht]
    \centering
    \label{QSpeckle}
    \caption{Proposed model's average scores on the testing dataset: (a) GT, (b) Input with speckle, (c) Sebastianelli et Al. \cite{sebastianelli2022speckle}
    and (d) Proposed \textbf{QSpeckleFilter.}\\}\label{tab:res}
    \begin{tabular}{lcc}
    \toprule
    Model & PSNR $\uparrow$ & SSIM $\uparrow$ \\
    \midrule
    (a) Ground Truth & $+\infty$ & 1.0 \\
    (b) Speckled & 15.70 & 0.58 \\
    \midrule
    (c) Sebastianelli et Al. \cite{sebastianelli2022speckle} & 19.21 & 0.75 \\
    (d) \textbf{QSPeckleFilter} & 21.72 & 0.81 \\
    \bottomrule
    \end{tabular}
\end{table}

\section{Discussion and Conclusions} \label{sec:disc_conc}
The conducted investigation into the realm of QML has revealed promising advancements in the domain of EO data processing. Through the utilization of quanvolutional layers, we have introduced the \textbf{Quanv4EO model}, a quanvolution method for preprocessing multi-dimensional EO data. Quanv4EO represents 
a novel approach to feature extraction, fully leveraging the power of quantum circuits to enhance classification tasks. Indeed, the training process, involving optimization of quantum circuits or variational parameters, presents computational challenges, yet the introduced methodology offers efficient access to high-dimensional Hilbert space, thus facilitating the manipulation of quantum states, measurements, and optimization algorithms.

The experiments conducted on both standard datasets like MNIST and FashionMNIST, as well as EO datasets such as EuroSAT, have demonstrated the effectiveness and efficiency of our proposed Quan4EO method. The results showcase high accuracy in image classification tasks, surpassing or matching SOTA models while maintaining significantly smaller model sizes. 

Moreover, this study extends beyond image classification, delving into unsupervised and supervised machine learning methodologies for EO data analysis. The application of BIRCH clustering, AutoML, and AutoDL to feature maps extracted by quanvolutional layers has yielded promising results, showcasing improved performance metrics compared to classical convolutional methods. Notably, the proposed approach demonstrates faster convergence rates and better classification accuracy across various land cover classes, further emphasizing its efficacy in handling EO data.

Looking ahead, the findings open avenues for future research, particularly in exploring the integration of quantum classifiers following the quanvolutional layers, potentially overcoming challenges associated with fully quantum approaches. Additionally, the prospect of training quanvolution filters presents opportunities for even more favorable results. Overall, this study contributes to advancing the field of quantum-enhanced machine learning and its applications in EO data analysis, paving the way for innovative solutions with real-world impact.\\
In the following, current limitations and possible solutions are highlighted: 
\paragraph{Current limitations and possible solutions} 

\begin{enumerate}
    \item \begin{itemize}
        \item \textbf{Problem} Number of features per quantum kernel/circuit is limited by the processing time. Kernel with more than 16 qubits already requires to much time to process 1 image.
        \item \textbf{Work around} Stack several smaller quantum kernels to have grater number of features maps
    \end{itemize}
    
    \item \begin{itemize}
        \item \textbf{Problem} Trainability of the circuit. To not lose the speed-up introduced by Jax, it is necessary to implement also the remaining part of the hybrid solution in Jax
        \item \textbf{Work around} Lazy training regime, it is a temporary solution we used as proof of concept, that already shows to achieve better results with respect to the counterpart
    \end{itemize}        
\end{enumerate}

Moreover, it is worth highlighting that different learning strategies have been tested and the proposed method achieved $2/5 \%$ of increase in performances. Since these results were obtained without training the quantum kernels, they are expected to increase. Moreover, the quantum classifier has proved to be more effective with respect to classic solutions and for this reason it is possible to expect that the combination of these variuous aspects will lead to a fully quantum solution able to overpass classical AI models.

\bibliographystyle{IEEEtran}
\bibliography{ref}

\end{document}